\documentclass
[preprint,prd,onecolumn,showpacs,showkeys,nobibnotes,nofootinbib,titlepage]{revtex4}%
\usepackage{amsfonts}
\usepackage{amsmath}
\usepackage{amssymb}
\usepackage{graphicx}
\usepackage{float}
\usepackage{rotating}%
\providecommand{\U}[1]{\protect\rule{.1in}{.1in}}

\ifx\pdfoutput\relax\let\pdfoutput=\undefined\fi
\newcount\msipdfoutput
\ifx\pdfoutput\undefined\else
\ifcase\pdfoutput\else
\ifx\paperwidth\undefined\else
\ifdim\paperheight=0pt\relax\else\pdfpageheight\paperheight\fi
\ifdim\paperwidth=0pt\relax\else\pdfpagewidth\paperwidth\fi
\fi\fi\fi
\begin{document}
\preprint{ }
\title[Wormholes in conformal Weyl gravity]{Wormhole geometries in fourth-order conformal Weyl gravity}
\author{Gabriele U. Varieschi}
\affiliation{Department of Physics, Loyola Marymount University - Los Angeles, CA 90045,
USA\footnote{Email: gvarieschi@lmu.edu}}
\eid{ }
\author{Kellie L. Ault}
\affiliation{Department of Physics, Loyola Marymount University - Los Angeles, CA 90045,
USA\footnote{Email: kault@lion.lmu.edu}}
\author{}
\affiliation{}

\begin{abstract}
We present an analysis of the classic wormhole geometries based on conformal
Weyl gravity, rather than standard general relativity. The main
characteristics of the resulting traversable wormholes remain the same as in
the seminal study by Morris and Thorne, namely, that effective super-luminal
motion is a viable consequence of the metric.

Improving on previous work on the subject, we show that for particular choices
of the shape and redshift functions the wormhole metric in the context of
conformal gravity does not violate the main energy conditions at or near the
wormhole throat.

Some exotic matter might still be needed at the junction between our solutions
and flat spacetime, but we demonstrate that the averaged null energy condition
(as evaluated along radial null geodesics) is satisfied for a particular set
of wormhole geometries.

Therefore, if fourth-order conformal Weyl gravity is a correct extension of
general relativity, traversable wormholes might become a realistic solution
for interstellar travel.

\end{abstract}
\startpage{1}
\endpage{ }
\maketitle
\tableofcontents

\section{\label{sect:introduction}Introduction}

Lorentzian wormholes are essentially shortcuts through space and time; they
can either connect two distant regions of our universe or connect our universe
with another, totally different one. In 1988 Morris and Thorne
\cite{1988AmJPh..56..395M}\ introduced a new class of \textit{traversable
wormholes}\ (TW, in the following) that, in principle, can be used for rapid
interstellar travel by advanced civilizations. Their solutions were entirely
based on standard General Relativity (GR) and represented a substantial
improvement over previous wormhole geometries, e.g., Schwarzschild wormholes,
Wheeler wormholes, and Kerr wormholes. For general reviews of Lorentzian
wormholes see Refs. \cite{1995lwet.book.....V} and \cite{Lobo:2007zb}.

However, even these improved TW solutions possess troublesome features
(\cite{1988AmJPh..56..395M}, \cite{1988PhRvL..61.1446M},
\cite{Friedman:2008dh}): they violate the main energy conditions for the
stress-energy tensor and they can also be converted into time machines, thus
possibly violating causality. The first of these two problems can only be
addressed by invoking the existence of exotic\ matter, i.e., matter with
negative energy density placed at or near the wormhole throat.\ Although some
special effects in quantum field theory (such as the Casimir effect, squeezed
vacuum, and others) might allow for the existence of exotic matter, the
manifest violation of the energy conditions greatly reduces the possibility to
use traversable wormholes for actual interstellar travel.

Notwithstanding these difficulties, progress has been made for wormholes in a
GR setting: attempts at self-consistently supporting wormholes with quantum
fields have been made (\cite{Hochberg:1996ee}, \cite{Butcher:2014lea}) and
classical fields as exotic matter have also been considered (see
\cite{Barcelo:1999hq}, \cite{Barcelo:2000zf}, and rebuttal
\cite{Butcher:2015sea}).

In recent decades, several alternative theories of gravity\ have been
introduced and applied to cosmological models \cite{Clifton:2011jh} with the
advantage of avoiding some of the most controversial elements of standard
cosmology---e.g., dark matter, dark energy, and the cosmological constant. In
particular, fourth-order Conformal Weyl Gravity (CG, for short, in the
following) is a natural extension of Einstein's GR, originally introduced by
H. Weyl \cite{Weyl:1918aa}, revisited by P. Mannheim et al.
(\cite{Mannheim:1988dj}, \cite{Kazanas:1988qa}, \cite{Mannheim:2005bfa},
\cite{2012FoPh...42..388M}), and even considered by G. 't Hooft
(\cite{Hooft:2014daa}, \cite{Hooft:2010ac}, \cite{Hooft:2010nc},
\cite{tHooft:2011aa}) as a possible key towards a complete understanding of
physics at the Planck scale.

A similar, but different approach to conformal gravity and cosmology has been
proposed by one of us in a series of papers (\cite{Varieschi:2008fc},
\cite{Varieschi:2008va}, \cite{Varieschi:2010xs}, \cite{Varieschi:2012ic},
\cite{Varieschi:2014ata}, \cite{2014Galax...2..577V}). A new model was
introduced that was called kinematical conformal cosmology
\cite{Varieschi:2008fc} since it was based on purely kinematic considerations
without using any dynamical equation of state for the universe. This model is
able to account for the accelerated expansion of the universe
(\cite{Varieschi:2008va}, \cite{2014Galax...2..577V}) and may also explain the
origin of some gravitational anomalies, such as the Flyby Anomaly
\cite{Varieschi:2014ata}.

One of these papers \cite{Varieschi:2012ic} studied in detail the implications
of CG in relation to the Alcubierre \textit{warp drive}\ metric
\cite{Alcubierre:1994tu}, which is another well-known theoretical mechanism
allowing, in principle, for super-luminal motion, i.e., faster-than-light
travel. It was shown \cite{Varieschi:2012ic} that, for particular choices of
the warp drive shaping function, the main energy conditions are not violated
by the warp drive solutions within the framework of conformal gravity.

Continuing along this line of investigation, the main objective of this paper
is to apply CG to traversable wormhole geometries and analyze the related
energy conditions and other characteristics of the resultant solutions. Some
work on the subject already exists in the literature
(\cite{2008CQGra..25q5006L}, \cite{2009IJMPA..24.1528O}), showing that in some
cases the energy conditions are not violated. We propose to find more general
wormhole solutions, which do not violate the energy conditions in CG, at least
in the vicinity of the wormhole throat. The same problem has been studied with
regard to other alternative theories of gravity (for example, see
\cite{Lemos:2003jb}, \cite{Boehmer:2007rm}, \cite{Lobo:2009ip},
\cite{2012AIPC.1458..447L}, \cite{Lobo:2012ai}, \cite{2013PhRvD..87f7504H}),
but in this paper we will limit our analysis to fourth-order conformal Weyl gravity.

In Sect. \ref{sect:conformal_gravity}, we start with a brief description of
the CG field equations and the general form of the CG stress-energy tensor; in
Sect. \ref{sect:traversable_wormholes},\ we review the standard TW geometry
and the related form of the stress energy tensor; in Sect.
\ref{sect:specific_cases}, we describe specific wormhole solutions and analyze
the main energy conditions, and, finally, in Sect. \ref{sect:conclusions}, we
present our conclusions.

\section{\label{sect:conformal_gravity}Conformal gravity and the stress-energy
tensor}

Conformal gravity is based on the Weyl action:\footnote{In this paper we adopt
a metric signature $(-,+,+,+)$ and we follow the sign conventions of Weinberg
\cite{Weinberg}. In this section we will leave fundamental constants, such as
$c$ and $G$, in all equations, but later we will use geometrized units ($c=1$,
$G=1$).}%

\begin{equation}
I_{W}=-\alpha_{g}\int d^{4}x\ (-g)^{1/2}\ C_{\lambda\mu\nu\kappa}%
\ C^{\lambda\mu\nu\kappa}, \label{eqn2.1}%
\end{equation}
where $g\equiv\det(g_{\mu\nu})$, $C_{\lambda\mu\nu\kappa}$ is the conformal
(or Weyl) tensor, and $\alpha_{g}$ is the CG coupling constant. $I_{W}$ is the
unique general coordinate scalar action that is invariant under local
conformal transformations: $g_{\mu\nu}(x)\rightarrow e^{2\alpha(x)}g_{\mu\nu
}(x)=\Omega^{2}(x)g_{\mu\nu}(x)$. The factor $\Omega(x)=e^{\alpha(x)}$
determines the amount of local stretching\ of the geometry, hence the name
conformal\ for a theory invariant under all local stretchings of the
space-time (see \cite{Varieschi:2008fc} and references therein for more details).

This conformally invariant generalization of GR was found to be a fourth-order
theory, as opposed to the standard second-order General Relativity, since the
field equations contained derivatives up to the fourth order of the metric
with respect to the space-time coordinates. These field equations were
introduced by R. Bach \cite{Bach:1921}, in the presence of a stress-energy
tensor\footnote{We follow here the convention \cite{Mannheim:2005bfa}\ of
introducing the stress-energy tensor $T_{\mu\nu}$ so that the quantity
$cT_{00}$ has the dimensions of an energy density.} $T_{\mu\nu}$, as follows:%

\begin{equation}
W_{\mu\nu}=\frac{1}{4\alpha_{g}}\ T_{\mu\nu}, \label{eqn2.2}%
\end{equation}
as opposed to Einstein's standard\ equations,%

\begin{equation}
G_{\mu\nu}=R_{\mu\nu}-\frac{1}{2}g_{\mu\nu}\ R=-\frac{8\pi G}{c^{3}}%
\ T_{\mu\nu}, \label{eqn2.3}%
\end{equation}
where the Bach tensor\ $W_{\mu\nu}$\ is the equivalent in CG of the Einstein
curvature tensor $G_{\mu\nu}$ on the left-hand side of Eq. (\ref{eqn2.3}).

$W_{\mu\nu}$ has a very complex structure and can be defined in a compact way as:%

\begin{equation}
W_{\mu\nu}=2C^{\alpha}{}_{\mu\nu}{}^{\beta}{}_{;\beta;\alpha}+C^{\alpha}%
{}_{\mu\nu}{}^{\beta}\ R_{\beta\alpha}, \label{eqn2.4}%
\end{equation}
or in an expanded form as:%

\begin{align}
W_{\mu\nu}  &  =-\frac{1}{6}g_{\mu\nu}\ R^{;\lambda}{}_{;\lambda}+\frac{2}%
{3}R_{;\mu;\nu}+R_{\mu\nu}{}^{;\lambda}{}_{;\lambda}-R_{\mu}{}^{\lambda}%
{}_{;\nu;\lambda}-R_{\nu}{}^{\lambda}{}_{;\mu;\lambda}+\frac{2}{3}R\ R_{\mu
\nu}\label{eqn2.5}\\
&  -2R_{\mu}{}^{\lambda}\ R_{\lambda\nu}+\frac{1}{2}g_{\mu\nu}\ R_{\lambda
\rho}\ R^{\lambda\rho}-\frac{1}{6}g_{\mu\nu}\ R^{2},\nonumber
\end{align}
involving derivatives up to the fourth order of the metric with respect to
space-time coordinates.

Therefore, in CG, the stress-energy tensor is computed by combining together
Eqs. (\ref{eqn2.2}) and (\ref{eqn2.5}):%

\begin{align}
T_{\mu\nu}  &  =4\alpha_{g}\ W_{\mu\nu}=4\alpha_{g}\ (-\frac{1}{6}g_{\mu\nu
}\ R^{;\lambda}{}_{;\lambda}+\frac{2}{3}R_{;\mu;\nu}+R_{\mu\nu}{}^{;\lambda}%
{}_{;\lambda}-R_{\mu}{}^{\lambda}{}_{;\nu;\lambda}-R_{\nu}{}^{\lambda}{}%
_{;\mu;\lambda}\label{eqn2.6}\\
&  +\frac{2}{3}R\ R_{\mu\nu}-2R_{\mu}{}^{\lambda}\ R_{\lambda\nu}+\frac{1}%
{2}g_{\mu\nu}\ R_{\lambda\rho}\ R^{\lambda\rho}-\frac{1}{6}g_{\mu\nu}%
\ R^{2}).\nonumber
\end{align}
This general form of the tensor will be used in the following sections, in
connection with the wormhole metric, to compute energy densities, pressures,
and other relevant quantities.

In the original papers by Mannheim and Kazanas (\cite{Mannheim:1988dj},
\cite{Kazanas:1988qa}), and in the CG\ wormhole analysis by F. Lobo
\cite{2008CQGra..25q5006L}, the computation of the non-zero components of
$W_{\mu\nu}$ was done indirectly by differentiating the Weyl action $I_{W}$
and using Bianchi and trace identities. In contrast, in this work, we compute
all the relevant tensors directly from their definitions, using a specialized
Mathematica program, which was developed by one of us for a previous study of
the warp drive in CG \cite{Varieschi:2012ic}.

In particular, this program can compute the conformal tensor $C_{\lambda\mu
\nu\kappa}$\ in Eq. (\ref{eqn2.1}), the Bach tensor\ $W_{\mu\nu}$ in Eq.
(\ref{eqn2.5}), and the stress-energy tensor $T_{\mu\nu}$\ in Eq.
(\ref{eqn2.6}), by performing all the necessary covariant derivatives. To
ensure the reliability of the results obtained with our program regarding
wormhole geometries, we have reproduced all the results discussed in the
CG\ wormhole analysis by F. Lobo \cite{2008CQGra..25q5006L}, obtaining a
perfect agreement.

\section{\label{sect:traversable_wormholes}Traversable wormholes and conformal
gravity}

\subsection{\label{subsect:general_form}General form of the metric}

The standard metric for traversable wormholes (\cite{1988AmJPh..56..395M},
\cite{1995lwet.book.....V}) can be written in two general forms. In terms of
the proper radial distance $l$ ($c=1$, in the following):%

\begin{equation}
ds^{2}=-e^{2\Phi(l)}dt^{2}+dl^{2}+r^{2}(l)\ \left[  d\theta^{2}+\sin^{2}%
\theta\ d\phi^{2}\right]  , \label{eqn3.1}%
\end{equation}
where $l$ covers the entire range $(-\infty,+\infty)$, $l=0$ corresponds to
the wormhole throat, and two asymptotically flat regions occur at $l=\pm
\infty$. Additional conditions \cite{1995lwet.book.....V} need to be obeyed by
the two functions $\Phi(l)$ and $r(l)$; in particular, $\Phi(l)$ needs to be
finite everywhere in order to avoid the existence of event horizons.

A more efficient way to express the TW metric is to employ Schwarzschild
coordinates ($t$, $r$, $\theta$, $\phi$):%

\begin{equation}
ds^{2}=-e^{2\Phi(r)}dt^{2}+\frac{dr^{2}}{1-b(r)/r}+r^{2}\ \left[  d\theta
^{2}+\sin^{2}\theta\ d\phi^{2}\right]  , \label{eqn3.2}%
\end{equation}
where now $\Phi(r)$ and $b(r)$ are two arbitrary functions of $r$,
respectively called the redshift\ and the shape\ function. The wormhole throat
corresponds to a minimum value of the radial coordinate, usually denoted by
$b_{0}$ or $r_{0}$ in the literature, so that two coordinate patches are now
required, each covering the range $[b_{0},+\infty)$, one for the upper
universe and one for the lower universe.\footnote{The redshift function
$\Phi(r)$ and the shape function $b(r)$ may actually be different on the upper
and lower portions of the wormhole (in this case, they are denoted by
$\Phi_{\pm}(r)$ and $b_{\pm}(r)$). In this paper, we will assume complete
symmetry between the upper and lower parts of the wormhole, thus we will need
only one function $\Phi(r)$ and only one function $b(r)$ over the range
$[b_{0},+\infty)$.}

The TW metric was originally introduced in 1973 by H. Ellis
\cite{Ellis:1973yv} as the \textit{drainhole}\ metric:%

\begin{equation}
ds^{2}=-dt^{2}+dl^{2}+r^{2}(l)\ \left[  d\theta^{2}+\sin^{2}\theta\ d\phi
^{2}\right]  , \label{eqn3.3}%
\end{equation}
corresponding to a special case of the general form in Eq. (\ref{eqn3.1}), for
$\Phi(l)=0$ and $r(l)=\sqrt{b_{0}^{2}+l^{2}}$. This metric was later
re-introduced and generalized by Morris-Thorne \cite{1988AmJPh..56..395M} in
1988, used as a tool for teaching general relativity by J. Hartle
\cite{2003gieg.book.....H}, visualized in various environments by M\"{u}ller
et al. \cite{Muller:2004dq}, and recently revisited by Thorne et al.
\cite{2015AmJPh..83..486J} after being popularized by the science fiction
movie \textit{Interstellar}.

The arbitrary functions $\Phi(r)$ and $b(r)$ in Eq. (\ref{eqn3.2}) also need
to satisfy certain conditions (see full details in \cite{1988AmJPh..56..395M},
or \cite{1995lwet.book.....V}) in order to obtain a viable TW solution. The
main condition for the redshift function is, again, the requirement for
$\Phi(r)$ to be finite everywhere to avoid the presence of event horizons.

The main conditions for the $b(r)$ function are related to the shape of the
wormhole, determined by the mathematics of embedding: we need $b(r=b_{0}%
)=b_{0}$ at the throat, $b(r)<r$ away from the throat, and the so-called
flaring-out\ condition \cite{1988AmJPh..56..395M},%

\begin{equation}
\frac{d^{2}r}{dz^{2}}=\frac{b(r)-b^{\prime}(r)\ r}{2b^{2}(r)}>0,
\label{eqn3.4}%
\end{equation}
at or near the throat. The function $r=r(z)$, or $z=z(r)$, determines the
profile of the embedding diagram of the wormhole:%

\begin{equation}
z(r)=\pm\int\nolimits_{b_{0}}^{r}\frac{dr}{\sqrt{\frac{r}{b(r)}-1}};
\label{eqn3.5}%
\end{equation}
the complete embedding diagram is obtained by rotating the graph of the
function $z(r)$ around the vertical $z$-axis.

\subsection{\label{subsect:energy_conditions}Energy conditions}

In a suitable orthonormal frame, the stress-energy tensor assumes a diagonal
form, $T^{\widehat{\mu}\widehat{\nu}}=diag(\rho,p_{1},p_{2},p_{3})$, where
$\rho$ is the energy density, and $p_{1}$, $p_{2}$, $p_{3}$ are the three
principal pressures.

In the case of the wormhole metric of Eq. (\ref{eqn3.2}), the spherical
symmetry of the metric implies $p_{2}=p_{3}$, so that the stress-energy tensor becomes%

\begin{equation}
T^{\widehat{\mu}\widehat{\nu}}=diag(\rho,p_{r},p_{l},p_{l}), \label{eqn3.6}%
\end{equation}
where $p_{r}$ indicates the pressure in the radial direction\footnote{The
radial tension $\tau$ is often used instead of the radial pressure $p_{r}$,
i.e., $\tau=-p_{r}$, especially when $\tau$ is a positive quantity, such as in
the case of wormholes in GR. In this paper, we prefer to use the radial
pressure $p_{r}$, since this quantity will be positive for most CG wormholes
analyzed in Sect. \ref{sect:specific_cases}.} and $p_{l}$ indicates the
pressure in the lateral ($\theta$ or $\phi$) directions.

The main energy conditions \cite{Hawking1} such as the null, weak, strong, and
dominant energy conditions (respectively, NEC, WEC, SEC, and DEC in the
following) for the stress-energy tensor in Eq. (\ref{eqn3.6}), can be
expressed directly in terms of $\rho$, $p_{r}$, $p_{l}$ as follows
(\cite{1995lwet.book.....V}, \cite{1984ucp..book.....W}):%

\begin{equation}%
\begin{array}
[c]{cc}%
NEC\Longleftrightarrow & \rho+p_{r}\geq0\text{ and }\rho+p_{l}\geq0\\
WEC\Longleftrightarrow & NEC\text{ and }\rho\geq0\\
SEC\Longleftrightarrow & NEC\text{ and }\rho+p_{r}+2p_{l}\geq0\\
DEC\Longleftrightarrow & \rho\geq\left\vert p_{r}\right\vert \text{ and }%
\rho\geq\left\vert p_{l}\right\vert .
\end{array}
\label{eqn3.7}%
\end{equation}

In the context of GR, the analysis of the traversable wormhole geometry has
shown (\cite{1988AmJPh..56..395M}, \cite{1995lwet.book.....V}) that the NEC is
violated over a finite range in the vicinity of the wormhole throat, thus
implying violation of the WEC, SEC, and DEC over the same range. At the throat
($r=b_{0}$), the weaker inequality $\left[  \rho(b_{0})+p_{r}(b_{0})\right]
\leq0$ holds, thus implying that the NEC is violated, or almost violated.
Moreover, the topological censorship theorem \cite{Friedman:1993ty} has shown
that GR\ does not allow an observer to probe the topology of spacetime, thus
implying that traversable wormholes require a violation of the averaged null
energy condition (ANEC). In any case, exotic matter, defined as violating
either the NEC or the WEC, must be threading the wormhole throat in all
possible cases in GR.

In CG, the situation might be different because of the different computation
of the stress-energy tensor using Eq. (\ref{eqn2.6}), instead of Eq.
(\ref{eqn2.3}). Also, as pointed out in Sect. II.B of Ref.
\cite{2008CQGra..25q5006L}, the CG analysis of the energy conditions might
require reconsidering the Raychaudhuri equation and its connection with the
topological censorship theorem \cite{Friedman:1993ty} and the ANEC
\cite{1995lwet.book.....V}.

It is beyond the scope of our current work to analyze the CG equivalent of the
topological censorship theorem and the role of the averaged energy conditions,
such as the ANEC. Therefore, in our current study, we will just take the
energy conditions in Eq. (\ref{eqn3.7}) at face value as was ultimately done
also in the same Ref. \cite{2008CQGra..25q5006L}. In addition, in Sect.
\ref{sect:specific_cases} we will simply check numerically the ANEC for the
geometries being considered.

As already mentioned in Sect. \ref{sect:conformal_gravity}, the rather
cumbersome computation of the stress-energy tensor has been performed by using
our specialized CG\ Mathematica program. Starting from the TW metric in Eq.
(\ref{eqn3.2}), all the relevant CG tensors were first evaluated in coordinate
basis, then transformed into the more convenient orthonormal basis.

In this basis, there is no difference between covariant and contravariant
forms of the stress-energy tensor; therefore, we have:%

\begin{align}
\rho &  =T^{\widehat{t}\widehat{t}}=T_{\widehat{t}\widehat{t}}\label{eqn3.8}\\
p_{r}  &  =T^{\widehat{r}\widehat{r}}=T_{\widehat{r}\widehat{r}}\nonumber\\
p_{l}  &  =T^{\widehat{\theta}\widehat{\theta}}=T_{\widehat{\theta
}\widehat{\theta}}=T^{\widehat{\phi}\widehat{\phi}}=T_{\widehat{\phi
}\widehat{\phi}}.\nonumber
\end{align}

The full expressions for $\rho$, $p_{r}$, and $p_{l}$, as a function of
$\Phi(r)$ and $b(r)$, are rather long and will be presented in the Appendix
(Sect. \ref{sect:appendix}). In the particular case of $\Phi(r)=0$ (the
so-called zero-tidal-force\ solutions \cite{1988AmJPh..56..395M}), the
simplified expressions are the following:%

\begin{align}
\rho &  =\frac{\alpha}{3r^{6}}\left\{  r^{2}\left[  -4r^{2}b^{(3)}%
+8rb^{\prime\prime}+b^{\prime}\left(  2rb^{\prime\prime}-5b^{\prime}-8\right)
\right]  +2rb\left[  9b^{\prime}+r\left(  2rb^{(3)}-5b^{\prime\prime}\right)
\right]  -9b^{2}\right\} \nonumber\\
p_{r}  &  =\frac{\alpha}{3r^{6}}\left\{  r^{2}\left[  -4rb^{\prime\prime
}-\left(  b^{\prime}-8\right)  b^{\prime}\right]  +2rb\left(  2rb^{\prime
\prime}-3b^{\prime}\right)  +3b^{2}\right\} \label{eqn3.9}\\
p_{l}  &  =\frac{\alpha}{3r^{6}}\left\{  r^{2}\left[  -2r^{2}b^{(3)}%
+6rb^{\prime\prime}+b^{\prime}\left(  rb^{\prime\prime}-2b^{\prime}-8\right)
\right]  +rb\left[  12b^{\prime}+r\left(  2rb^{(3)}-7b^{\prime\prime}\right)
\right]  -6b^{2}\right\}  ,\nonumber
\end{align}
where the first two expressions for $\rho$ and $p_{r}$ are equivalent to those
of Eqs. (28)-(29) in Ref. \cite{2008CQGra..25q5006L}.\footnote{In Ref.
\cite{2008CQGra..25q5006L}, the quantity $4\alpha_{g}$, from Eq.
(\ref{eqn2.6}), is set to one. In this paper, we prefer to leave this quantity
in all our general formulas, such as Eqs. (\ref{eqn3.9}), and (\ref{eqnA.1}%
)-(\ref{eqnA.3}).}

\section{\label{sect:specific_cases}Specific cases and related energy
conditions}

In the previous sections, we discussed at length the general connections
between TW geometries and conformal Weyl gravity. In this section, we analyze
specific cases for different choices of the functions $\Phi(r)$ and $b(r)$,
with the aim of finding solutions which do not violate the energy conditions
in the vicinity of the throat.

In general, different strategies can be used to find suitable functions
$\Phi(r)$ and $b(r)$: one can start by choosing simple algebraic expressions
for $\Phi$ and $b$, as was done for example in Refs.
\cite{1988AmJPh..56..395M} and \cite{2008CQGra..25q5006L}, which lead to
wormholes with the desired geometries. In this case, one must also make sure
that the chosen functions satisfy the additional conditions mentioned in Sect.
\ref{subsect:general_form}, e.g., the flaring-out condition in Eq.
(\ref{eqn3.4}).

An alternative strategy is to start with a well-designed shape of the
wormhole, given by the profile of the embedding diagram $z=z(r)$, and then use
Eq. (\ref{eqn3.5}) to determine the related shape function $b(r)$. This was
the case, for example, of the recent study by Thorne et al.
\cite{2015AmJPh..83..486J}, where a particular geometry with three free
shaping parameters (Dneg---Double Negative wormhole) was used to visualize the wormhole.

In this work, we prefer the former approach to the choice of the shape and
redshift functions. In an effort to generalize the different forms for $b$ The
above figureand $\Phi$ existing in the literature, we considered the following functions:%

\begin{align}
b(r)  &  =b_{0}\left[  \left(  1+\gamma\right)  \left(  \frac{b_{0}}%
{r}\right)  ^{l}-\gamma\left(  \frac{b_{0}}{r}\right)  ^{m}\right]
\label{eqn4.1}\\
\Phi(r)  &  =\left[  \left(  1-\delta\right)  \left(  \frac{b_{0}}{r}\right)
^{p}-\delta\left(  \frac{b_{0}}{r}\right)  ^{q}\right]  ,\nonumber
\end{align}
where $\gamma$, $l$, $m$, $\delta$, $p$, and $q$ can be considered as free parameters.

Both functions are made of simple power terms of the ratio $b_{0}/r$, similar
to those used in the cited literature (\cite{1988AmJPh..56..395M},
\cite{2008CQGra..25q5006L}). The $b(r)$ function correctly reduces to $b_{0}$
for $r=b_{0}$, and the $\Phi(r)$ function is finite for $r\geq b_{0}$. The
energy conditions in Eq. (\ref{eqn3.7}) and the flaring-out condition in Eq.
(\ref{eqn3.4}) will be satisfied for certain particular ranges of values of
the six free parameters in the CG\ case.

Our Mathematica program allowed us to find these particular values for the
free parameters, satisfying the energy conditions in CG. We combined the
general CG equations shown in the Appendix for $\rho$, $p_{r}$, and $p_{l}$
(Eqs. \ref{eqnA.1}-\ref{eqnA.3}) together with the expressions in Eq.
(\ref{eqn4.1}). Then, we computed all the quantities in the energy conditions
of Eq. (\ref{eqn3.7}) plus the flaring-out condition of Eq. (\ref{eqn3.4}).
The results thus obtained were plotted as a function of the six free
parameters and several solutions were found, satisfying the energy conditions
at and near the throat. We present some of these solutions in the following subsections.

\subsection{\label{subsect:zero_solutions}Solutions with $\Phi=0$}

The simplest class of solutions is obtained by setting the redshift function
to zero, i.e., $\Phi(r)=0$. As already mentioned above, these are also called
zero-tidal-force solutions \cite{1988AmJPh..56..395M} because no radial tidal
acceleration is felt by travelers crossing the wormhole. This class of
solutions is also obtained by setting $p=q=0$ and $\delta=1/2$ in the second
line of Eq. (\ref{eqn4.1}) above. In this case, we are left with the choice of
the three free parameters, $\gamma$, $l$, and $m$, for the shape function
$b(r)$ in the first line of Eq. (\ref{eqn4.1}).

By setting $\gamma=0$ or $\gamma=-1$, the function $b(r)$ reduces to a single
power term. Inspection of the graphs of the energy and flaring-out conditions
shows that no valid solutions exist in this case: either the NEC is violated,
or the flaring-out condition is violated, or both. The same happens for values
of $\gamma$ in the range $-1<\gamma<0$.

In contrast, valid solutions can be obtained for $\gamma>0$, or $\gamma<-1$,
by adjusting the values of the remaining parameters $l$ and $m$. Since all
these solutions are similar to each other, in Fig. 1 we present one particular
example for $\gamma=1$, $l=3$, and $m=6.5$, which illustrates well these types
of solutions. However, we remark that infinitely many other valid solutions
can be obtained by adjusting the free parameters within the allowed ranges;
the case presented here is by no means unique.

In Fig. 1, we show all the functions related to the energy conditions in Eq.
(\ref{eqn3.7}) and the flaring-out condition in Eq. (\ref{eqn3.4}) for the
$b(r)$ function with parameters $\gamma=1$, $l=3$, and $m=6.5$ (for $\Phi=0$).
In this figure, we also set $b_{0}=1$, so that $r\geq1$, and the (currently
unknown) value of the conformal coupling constant $\alpha$ is also set to unity.

We can see immediately that all the plotted energy functions\ are positive at
and near the throat (in the range $1\leq r\lesssim1.03$), as evidenced also by
the shaded areas (light green---positive; light red---negative); in addition,
the flaring-out condition is also satisfied for all values or $r$. Therefore,
for this case, all the main energy conditions---NEC, WEC, SEC, and DEC--- are
satisfied in the vicinity of the throat, and no exotic matter is needed in
this region, as opposed to the GR case.

The wormhole geometry related to this solution will need to be matched to
either a flat spacetime, or to the standard CG vacuum solution
\cite{Mannheim:1988dj}, at about $r\simeq1.03$ in order to avoid exotic matter
beyond this value of the radial coordinate. This can be accomplished by
applying appropriate junction conditions\ as in the case of GR
\cite{1973grav.book.....M}.

However, when we tried to match our solutions with $\Phi(r)=0$ to flat
spacetime in several different ways using our Mathematica routines (with
either a discontinuous or a smooth transition of the $b(r)$ function into flat
spacetime at $r=1.03$), it was found that some exotic matter needed to be
present at the junction. At this point, we are unable to ascertain whether the
presence of exotic matter at the junction can be avoided by using a different
shape function $b(r)$ or if this is a general feature of all wormhole
geometries in CG.

In relation to the already mentioned topological censorship theorem
\cite{Friedman:1993ty} and the violation in GR of the averaged null energy
condition, it is interesting to evaluate the ANEC for our wormhole geometries,
at least for radial null geodesics denoted by $\Gamma$ in the following. For
this case, it is sufficient to evaluate the ANEC\ integral (see Ref.
\cite{1995lwet.book.....V}, Sect. 12.4.2 and Eq. 12.63):%

\begin{equation}
I_{\Gamma}=\int_{l=-\infty}^{l=+\infty}\left[  \rho(l)-\tau(l)\right]
e^{-\Phi(l)}dl=2\int_{r=b_{0}}^{r=\infty}\frac{\left[  \rho(r)+p_{r}%
(r)\right]  e^{-\Phi(r)}}{\sqrt{1-b(r)/r}}dr. \label{eqn4.2}%
\end{equation}
This integral is first expressed as a function of the proper radial distance
$l$ used in the metric of Eq. (\ref{eqn3.1}), then converted into the radial
coordinate $r$ by using the connection $dl=\pm\frac{dr}{\sqrt{1-b(r)/r}}$, for
the upper and lower parts of the wormhole respectively. We also used the
relation $\tau=-p_{r}$, see footnote after Eq. (\ref{eqn3.6}).

We computed numerically the ANEC integral in Eq. (\ref{eqn4.2}) for the class
of $\Phi=0$ solutions considered in this subsection for different values of
the parameters $\gamma$, $l$, and $m$ in Eq. (\ref{eqn4.1}), always obtaining
negative results. Thus, the ANEC is violated for specific cases of this class
of wormhole geometries (in particular, for $b_{0}=1$, $\alpha=1$, $\gamma=1$,
$l=3$, and $m=6.5$ we obtained $I_{\Gamma}=-6.35\
\operatorname{cm}%
^{-1}$). This suggests that the violation is a generic feature of this class
of geometries. In contrast, positive results of the ANEC integral will be
obtained for the class of $\Phi\neq0$ solutions described in the next
subsection \ref{subsect:nonzero_solutions}.

In summary, our solutions for the $\Phi=0$ case, represent a generalization
and a definite improvement over the solutions presented in Sect. III. A. of
Ref. \cite{2008CQGra..25q5006L}. In this reference, the CG solutions for
$\Phi=0$ were either violating all the energy conditions, or were satisfying
only the NEC while violating the other conditions. In contrast, we have shown
that in CG,\ it is easy to find a class of solutions, even in the $\Phi=0$
case, which does not violate any energy condition at or near the throat.

In addition to the solutions presented above, we also tried several other
shape functions different from the one described in Eq. (\ref{eqn4.1}). Adding
other power terms to the function $b(r)$ does not significantly change the
outcome of the analysis: for certain values of the free parameters we obtain
valid solutions similar to the ones already described. Starting instead with
predefined shapes for the wormhole (such as the Ellis wormhole, the Dneg
wormhole, or others), does not yield solutions in CG which satisfy completely
the energy conditions; therefore, we will not consider here these other solutions.

\subsection{\label{subsect:nonzero_solutions}Solutions with $\Phi\neq0$}

Another class of solutions can be obtained by setting $\Phi\neq0$, using the
special form for $\Phi$ in Eq. (\ref{eqn4.1}). Again, this particular $\Phi
$\ function generalizes other redshift functions used in the literature, and
we adopted a combination of two power terms of the ratio $b_{0}/r$ in order to
keep this function as simple as possible.

It should be noted that for $\delta=0$ in Eq. (\ref{eqn4.1}), the redshift
function reduces to $\Phi(r)=\left(  b_{0}/r\right)  ^{p}$, similar to the one
used in Sect. III. B. of Ref. \cite{2008CQGra..25q5006L} (with $p=1$), while
for $\delta=1$ the redshift function reduces to $-\left(  b_{0}/r\right)
^{q}$, similar to the one used in Sect. III. C. of the same cited reference
(with $q=1$). Our more general choice for $\Phi(r)$ allows us to change the
values of the free parameters $\delta$, $p$, and $q$, in order to explore all
possible cases.

However, after extensive testing of all possible combinations of the
parameters, we found that the $\Phi\neq0$ solutions are not much different
from the $\Phi=0$ solutions studied in the previous subsection, at least
energetically. Depending on the values of the six free parameters, we either
obtain solutions which satisfy the energy condition up to a certain maximum
value for $r$, or solutions which clearly violate them. The presence of the
$\Phi(r)$ term in the metric simply adds some sort of fine-tuning\ effect to
the $\Phi=0$ solutions, but does not change significantly the analysis of the
energy conditions.

For example, Fig. 2 shows one particular solution for the following parameter
values: $\delta=1$, $q=1$, $\gamma=1$, $l=3$, and $m=5$ (also, $b_{0}=1$ and
$\alpha=1$). Therefore, it is a special case with $\Phi(r)=-\left(
b_{0}/r\right)  $, similar to the one in Sect. III. C. of Ref.
\cite{2008CQGra..25q5006L}. Once again, the plotted energy functions clearly
show that all energy conditions are satisfied in the vicinity of the throat,
for $1\leq r\lesssim1.10,$ and that the flaring-out condition is also
satisfied over the whole range of $r$.

As already discussed in Sect. \ref{subsect:zero_solutions} for the $\Phi=0$
case, we also matched our $\Phi\neq0$ solutions described above with flat
spacetime at $r=1.10$ with similar results: some exotic matter is needed at
the junction also in the $\Phi\neq0$ case.

However, the analysis of the ANEC integral in Eq. (\ref{eqn4.2}) yields
completely different results in this case. Due to the presence of the
$e^{-\Phi(r)}$ term in this equation (with $\Phi\neq0$, in this case), the
(numerically integrated) value of the ANEC condition is now \emph{positive}
for a wide range of the parameters. In particular, for $b_{0}=1$, $\alpha=1$,
$\delta=1$, $q=1$, $\gamma=1$, $l=3$, and $m=5$ (corresponding to the solution
presented in Fig. 2), we obtained $I_{\Gamma}=5.99\
\operatorname{cm}%
^{-1}$. Positive values of the ANEC integral are also obtained if we match our
$\Phi\neq0$ solutions with flat spacetime at $r\simeq1.10$, with smooth or
discontinuous transitions.

Although the ANEC\ integrals were computed just along radial null geodesics,
the positive values obtained in this subsection seem to indicate that the
averaged null energy condition is satisfied for our $\Phi\neq0$ solutions.
This suggests that the topological censorship theorem may not apply to
conformal gravity.

In figures 3 and 4, we show the actual shape of the wormholes studied above by
plotting the standard embedding diagrams for $b_{0}=1$, with the other
parameters set as before. These embedding diagrams were obtained by rotating
the graph of the function $z(r)$ in Eq. (\ref{eqn3.5}) around the vertical
$z$-axis. The correct shape of the wormholes, illustrated in Figs. 3-4,
confirms that the flaring-out condition is verified for these solutions.

\section{\label{sect:conclusions}Conclusions}

In this paper, we have analyzed in detail different wormhole geometries within
the framework of fourth-order Conformal Weyl Gravity. We have seen that, for
particular choices of the shape and redshift functions, we can overcome the
main limitation of TW in standard General Relativity, namely, the violation of
the energy conditions at or near the throat.

In fact, we have shown that for several different wormhole shapes, the CG
solutions do not violate any energy condition in the vicinity of the throat.
In particular, this can be achieved even with $\Phi=0$, zero-tidal-force
solutions, unlike previous works in the literature that have only found viable
$\Phi\neq0$ solutions.

Some exotic matter might still be needed at the junction between our solutions
and flat spacetime, although more work will be needed to analyze in detail the
junction condition formalism in CG. Similarly, the connection between the
Raychaudhuri equation, the topological censorship theorem, and the ANEC might
need to be reconsidered in CG. Our analysis of the ANEC has shown that this
condition is likely to be satisfied for our class of $\Phi\neq0$ wormholes in
CG, thus implying that it is qualitatively easier to make traversable
wormholes in conformal gravity than in GR.

The TW mechanism might be theoretically feasible if CG is the correct
extension of current gravitational theories. All components of the
stress-energy tensor can be analytically calculated, using our specialized
Mathematica program based on Conformal Gravity. Therefore, traversable
wormholes can be, at least in principle, fully engineered following our computations.

\section{\label{sect:appendix}Appendix: energy density and pressure
expressions in conformal gravity}

We present here the general expressions for the energy density $\rho$, the
radial pressure $p_{r}$, and the lateral pressure $p_{l}$ in Conformal
Gravity, computed using our specialized Mathematica program from the metric in
Eq. (\ref{eqn3.2}). These quantities are defined in terms of the arbitrary
functions $\Phi(r)$ and $b(r)$, and their derivatives, as follows:%

\begin{align}
\rho &  =\frac{\alpha}{3r^{6}}\{r^{2}[-b^{\prime}{}^{2}(r\left(
6r\Phi^{\prime\prime}+\Phi^{\prime}\left(  3r\Phi^{\prime}-2\right)  \right)
+5)+4r(2b^{\prime\prime}+r(r\Phi^{\prime}(b^{(3)}-4r\Phi^{(3)}-22\Phi
^{\prime\prime})\nonumber\\
&  -b^{(3)}+\Phi^{\prime}{}^{2}\left(  2r\left(  b^{\prime\prime}%
+r\Phi^{\prime\prime}\right)  -5\right)  -r(-4b^{\prime\prime}\Phi
^{\prime\prime}+2r\Phi^{(4)}+8\Phi^{(3)}+3r\Phi^{\prime\prime}{}^{2}%
)+r^{2}\Phi^{\prime}{}^{4}-2r\Phi^{\prime}{}^{3}))\nonumber\\
&  +b^{\prime}(2r(\Phi^{\prime}\left(  -rb^{\prime\prime}+18r^{2}\Phi
^{\prime\prime}-10\right)  +b^{\prime\prime}-2r^{2}\Phi^{\prime}{}^{3}%
+20r\Phi^{\prime}{}^{2}+2r(6r\Phi^{(3)}+5\Phi^{\prime\prime}))-8)]\nonumber\\
&  -2rb[b^{\prime}(r(-2r^{2}\Phi^{\prime}{}^{3}+2\Phi^{\prime}\left(
9r^{2}\Phi^{\prime\prime}-4\right)  +17r\Phi^{\prime}{}^{2}+4r(3r\Phi
^{(3)}+\Phi^{\prime\prime}))-9)\tag{A.1}\label{eqnA.1}\\
&  +r(4r^{2}\Phi^{\prime}{}^{2}\left(  b^{\prime\prime}+2r\Phi^{\prime\prime
}\right)  +5b^{\prime\prime}-2r(b^{(3)}-\left(  4rb^{\prime\prime}+5\right)
\Phi^{\prime\prime}+4r^{2}\Phi^{(4)}+6r^{2}\Phi^{\prime\prime}{}^{2}%
+10r\Phi^{(3)})\nonumber\\
&  -\Phi^{\prime}(r(2r(-b^{(3)}+8r\Phi^{(3)}+35\Phi^{\prime\prime}%
)+b^{\prime\prime})+10)+4r^{3}\Phi^{\prime}{}^{4}-10r^{2}\Phi^{\prime}{}%
^{3})]\nonumber\\
&  +b^{2}[r(4r^{3}\Phi^{\prime}{}^{4}-12r^{2}\Phi^{\prime}{}^{3}+r\Phi
^{\prime}{}^{2}\left(  8r^{2}\Phi^{\prime\prime}+17\right)  -2r(6r^{2}%
\Phi^{\prime\prime}{}^{2}-7\Phi^{\prime\prime}+4r(r\Phi^{(4)}+\Phi
^{(3)}))\nonumber\\
&  -2\Phi^{\prime}(8r^{3}\Phi^{(3)}+26r^{2}\Phi^{\prime\prime}%
+9))-9]\}\nonumber
\end{align}

\begin{align}
p_{r}  &  =\frac{\alpha}{3r^{6}}\{8r^{4}\Phi^{\prime\prime}[b^{\prime}\left(
1-r\Phi^{\prime}\right)  +r^{2}\Phi^{\prime}{}^{2}-2]\tag{A.2}\label{eqnA.2}\\
&  -r^{2}\left(  r\Phi^{\prime}-1\right)  [4r\left(  r\Phi^{\prime}-1\right)
\left(  b^{\prime\prime}+\Phi^{\prime}\left(  r\Phi^{\prime}-4\right)
\right)  +b^{\prime}{}^{2}\left(  r\Phi^{\prime}-1\right)  +4b^{\prime}\left(
r\Phi^{\prime}\left(  r\Phi^{\prime}-1\right)  +2\right)  ]\nonumber\\
&  +2rb[2r(-\Phi^{\prime}(2r(b^{\prime\prime}+2r^{2}\Phi^{(3)}-r\Phi
^{\prime\prime})+5)+r\Phi^{\prime}{}^{2}\left(  r\left(  b^{\prime\prime
}-4r\Phi^{\prime\prime}\right)  +16\right)  +b^{\prime\prime}+2r^{3}%
\Phi^{\prime}{}^{4}\nonumber\\
&  -11r^{2}\Phi^{\prime}{}^{3}+2r(\Phi^{\prime\prime}\left(  r^{2}\Phi
^{\prime\prime}+3\right)  +2r\Phi^{(3)}))+b^{\prime}\left(  r\Phi^{\prime
}-1\right)  \left(  r\left(  4r\Phi^{\prime\prime}+\Phi^{\prime}\left(
2r\Phi^{\prime}-1\right)  \right)  +3\right)  ]\nonumber\\
&  +b^{2}[r(-4r^{3}\Phi^{\prime}{}^{4}+20r^{2}\Phi^{\prime}{}^{3}%
-4r(\Phi^{\prime\prime}\left(  r^{2}\Phi^{\prime\prime}+2\right)
+2r\Phi^{(3)})+r\Phi^{\prime}{}^{2}\left(  8r^{2}\Phi^{\prime\prime}-29\right)
\nonumber\\
&  +\Phi^{\prime}(8r^{3}\Phi^{(3)}-8r^{2}\Phi^{\prime\prime}+6))+3]-4r^{6}%
\Phi^{\prime\prime}{}^{2}+8r^{5}\Phi^{(3)}\left(  r\Phi^{\prime}-1\right)
\}\nonumber
\end{align}

\begin{align}
p_{l}  &  =\frac{\alpha}{3r^{6}}\{r^{2}[b^{\prime}{}^{2}\left(  -r^{2}\left(
3\Phi^{\prime\prime}+\Phi^{\prime}{}^{2}\right)  -2\right)  +2r(3b^{\prime
\prime}+r(-b^{(3)}+\left(  3rb^{\prime\prime}+4\right)  \Phi^{\prime}{}%
^{2}\tag{A.3}\label{eqnA.3}\\
&  -2r(-2b^{\prime\prime}\Phi^{\prime\prime}+3\Phi^{(3)}+r(\Phi^{(4)}%
+\Phi^{\prime\prime}{}^{2}))+\Phi^{\prime}(r(b^{(3)}-6r\Phi^{(3)}%
-22\Phi^{\prime\prime})-2b^{\prime\prime})+2r^{2}\Phi^{\prime}{}^{4}%
+4\Phi^{\prime\prime}\nonumber\\
&  -8r\Phi^{\prime}{}^{3}))+b^{\prime}(r(\Phi^{\prime}\left(  -rb^{\prime
\prime}+22r^{2}\Phi^{\prime\prime}-4\right)  +b^{\prime\prime}+16r\Phi
^{\prime}{}^{2}+6r(2r\Phi^{(3)}+\Phi^{\prime\prime}))-8)-8r\Phi^{\prime
}]\nonumber\\
&  +rb[r(-2r\left(  3rb^{\prime\prime}+16\right)  \Phi^{\prime}{}%
^{2}-7b^{\prime\prime}+2r(b^{(3)}-\left(  4rb^{\prime\prime}+11\right)
\Phi^{\prime\prime}+4r^{2}\Phi^{(4)}+4r^{2}\Phi^{\prime\prime}{}^{2}%
+6r\Phi^{(3)})\nonumber\\
&  +\Phi^{\prime}(2r^{2}(-b^{(3)}+12r\Phi^{(3)}+33\Phi^{\prime\prime
})+5rb^{\prime\prime}+20)-8r^{3}\Phi^{\prime}{}^{4}+32r^{2}\Phi^{\prime}{}%
^{3})-2b^{\prime}(6r^{3}\Phi^{(3)}\nonumber\\
&  +r\Phi^{\prime}\left(  11r^{2}\Phi^{\prime\prime}+7r\Phi^{\prime}-2\right)
-6)]+b^{2}[r(-4r^{3}\Phi^{(4)}-4r^{3}\Phi^{\prime\prime}{}^{2}+4r^{3}%
\Phi^{\prime}{}^{4}-16r^{2}\Phi^{\prime}{}^{3}\nonumber\\
&  -2\Phi^{\prime}(6r^{3}\Phi^{(3)}+11r^{2}\Phi^{\prime\prime}+6)+11r\Phi
^{\prime\prime}+23r\Phi^{\prime}{}^{2})-6]\}\nonumber
\end{align}

\begin{acknowledgments}
This work was supported by a grant from the Frank R. Seaver College of Science
and Engineering, Loyola Marymount University. The authors would like to thank
the anonymous reviewer for the valuable suggestions and useful comments which
helped improve the final version of this paper.
\end{acknowledgments}

\bibliographystyle{apsrev}
\bibliography{WORMHOLES}
\newpage

\begin{center}
\begin{figure}[ptb]
\centering
\ifcase\msipdfoutput
\includegraphics[
width=\textwidth
]
{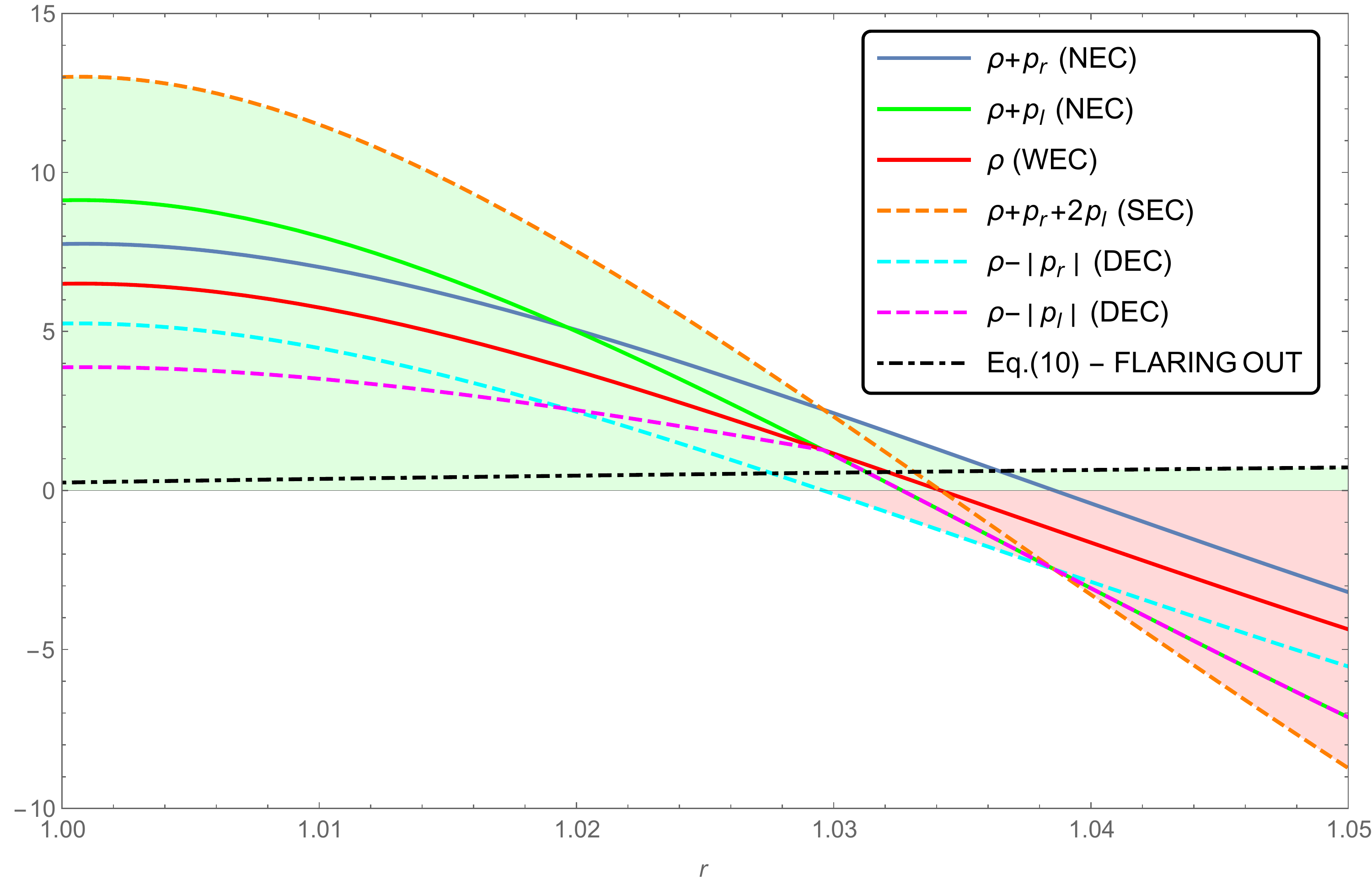}\else
\includegraphics[
width=\textwidth
]
{figure1.pdf}\fi
\label{fig1}\end{figure}
\end{center}

Figure 1: The functions related to the energy conditions in Eq. (\ref{eqn3.7})
and the flaring-out condition in Eq. (\ref{eqn3.4}) are shown here for a
particular shape function with parameters $\gamma=1$, $l=3$, and $m=6.5$ (for
$\Phi=0$). We also set $b_{0}=1$ and $\alpha=1$. Shaded areas (light
green---positive; light red---negative) are used to emphasize regions where
the energy conditions are satisfied or violated. All plotted energy
functions\ are positive at and near the throat (in the range $1\leq
r\lesssim1.03$) showing that no exotic matter is needed in CG as opposed to
the GR case.

\newpage

\begin{figure}[ptb]
\centering
\ifcase\msipdfoutput
\includegraphics[
width=\textwidth
]
{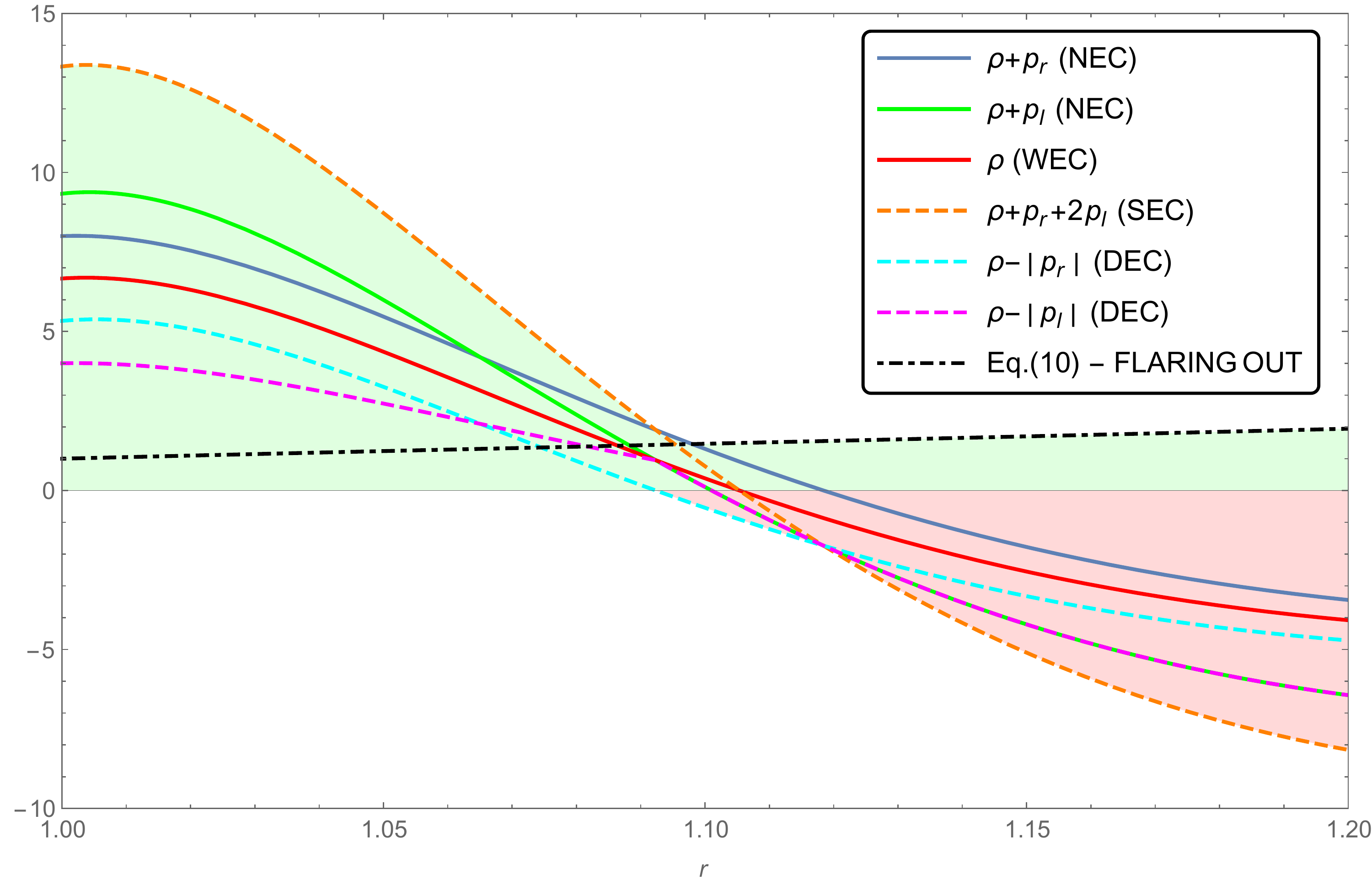}\else
\includegraphics[
width=\textwidth
]
{figure2.pdf}\fi
\label{fig2}\end{figure}

Figure 2: The functions related to the energy conditions in Eq. (\ref{eqn3.7})
and the flaring-out condition in Eq. (\ref{eqn3.4}) are shown here for a
particular shape function with parameters $\delta=1$, $q=1$, $\gamma=1$,
$l=3$, and $m=5$ ($\Phi\neq0$ case). We also set $b_{0}=1$ and $\alpha=1$.
Shaded areas (light green---positive; light red---negative) are used to
emphasize regions where the energy conditions are satisfied or violated. All
plotted energy functions\ are positive at and near the throat (in the range
$1\leq r\lesssim1.10$) showing that no exotic matter is needed in CG as
opposed to the GR case.

\newpage

\begin{figure}[ptb]
\centering
\ifcase\msipdfoutput
\includegraphics[
width=\textwidth
]
{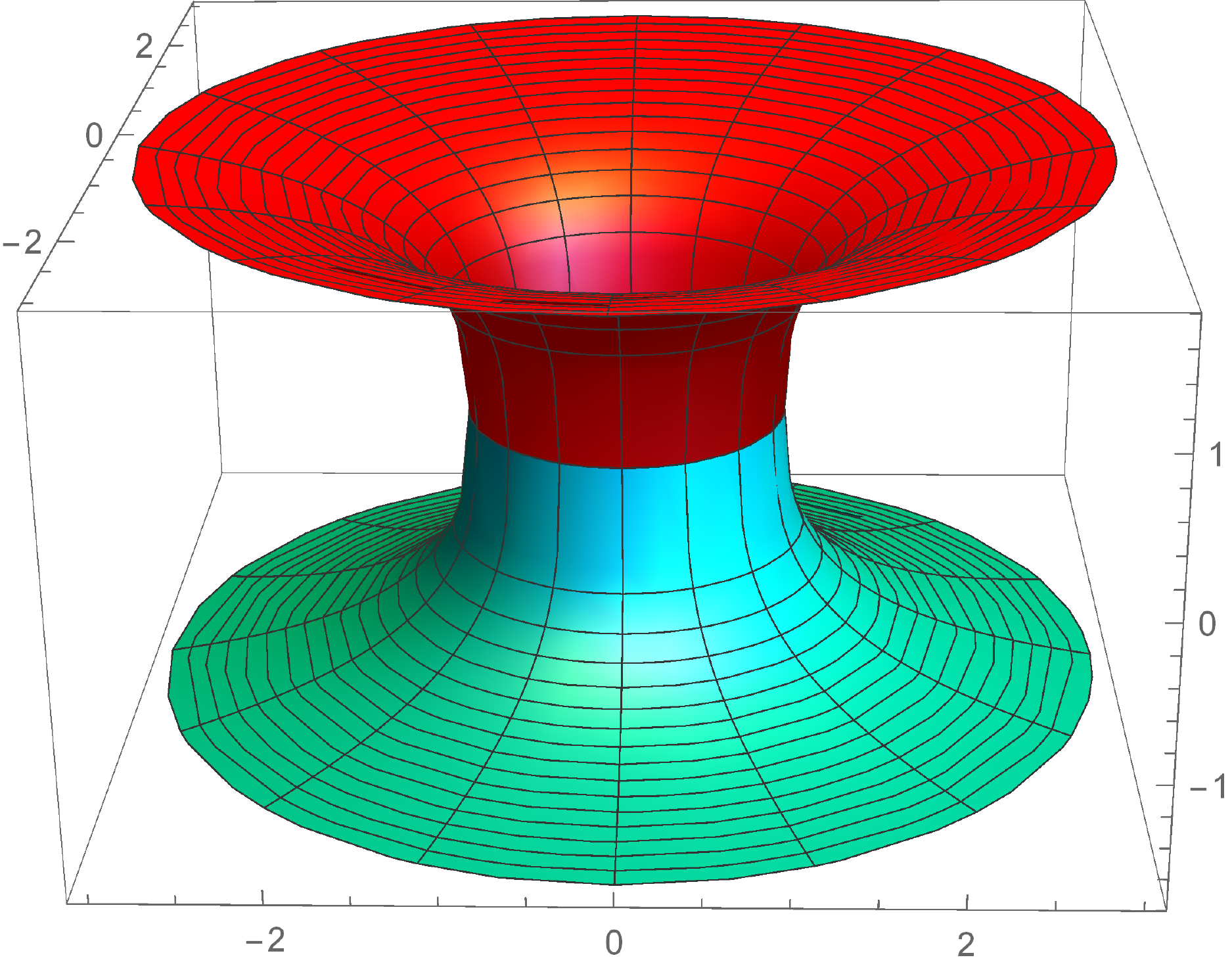}\else
\includegraphics[
width=\textwidth
]
{figure3.pdf}\fi
\label{fig3}\end{figure}Figure 3: The shape of the wormhole studied in Fig. 1
is illustrated here by plotting the standard embedding diagram for $b_{0}=1$,
$\alpha=1$, and with the other parameters set as before. The shape of the
wormhole confirms that the flaring-out condition is verified for this solution
(this figure was not included in the published version of the paper).

\newpage

\begin{figure}[ptb]
\centering
\ifcase\msipdfoutput
\includegraphics[
width=\textwidth
]
{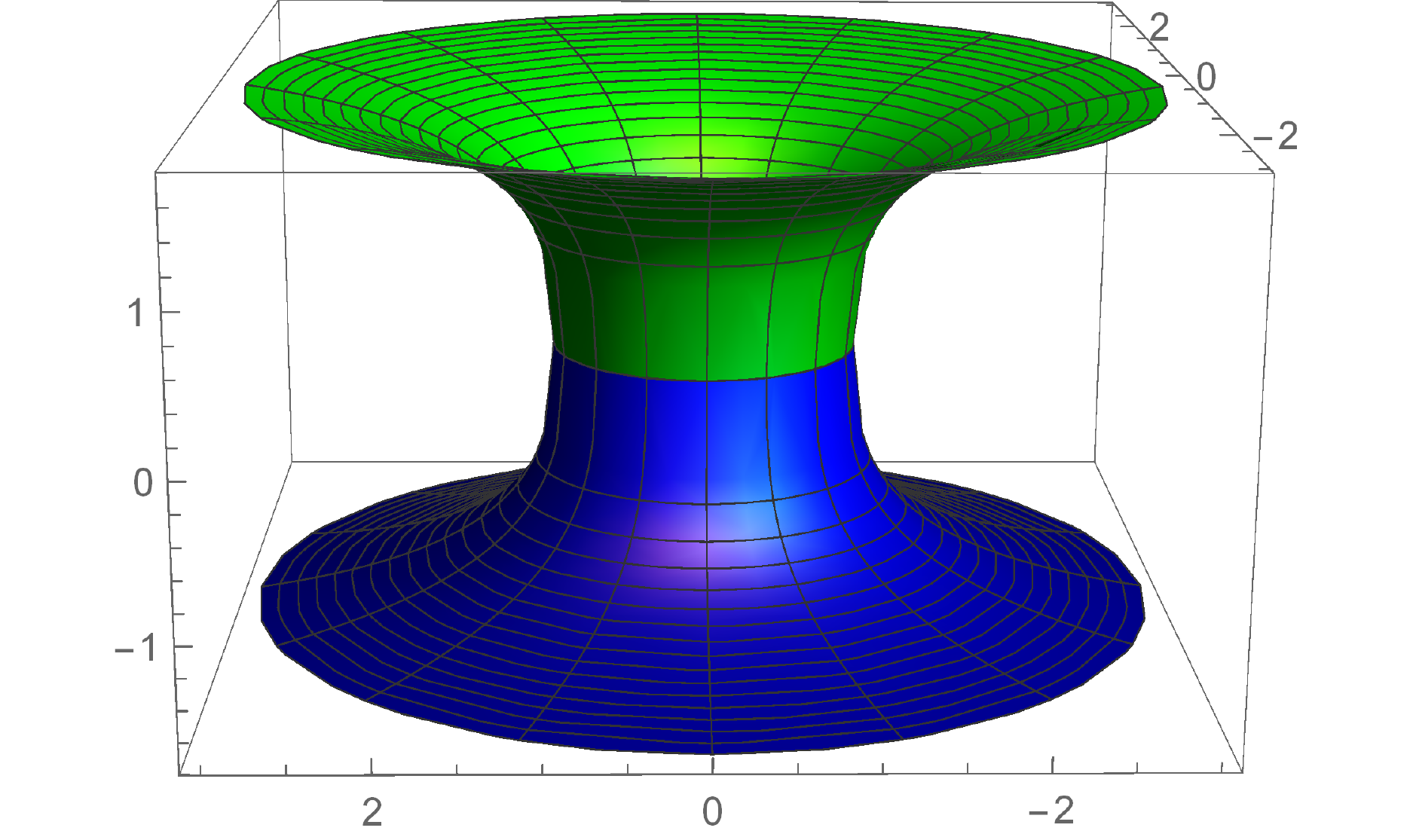}\else
\includegraphics[
width=\textwidth
]
{figure4.pdf}\fi
\label{fig4}\end{figure}Figure 4: The shape of the wormhole studied in Fig. 2
is illustrated here by plotting the standard embedding diagram for $b_{0}=1$,
$\alpha=1$, and with the other parameters set as before. The shape of the
wormhole confirms that the flaring-out condition is verified for this solution
(this figure was not included in the published version of the paper).
\end{document}